\documentclass[a4paper,UKenglish]{lipics-v2016}
\usepackage{microtype}%if unwanted, comment out or use option "draft"

\title{Skyline Queries in O(1) time?}
\author[1]{S. Sioutas}
\author[2]{K. Tsichlas}
\author[2]{A. Kosmatopoulos}
\author[2]{A. Papadopoulos}
\author[1]{D. Tsoumakos}
\author[3]{K. Doka}
\affil[1]{Ionian University, Information Systems and Databases Laboratory, Corfu, Greece\\
  \texttt{sioutas@ionio.gr}, \texttt{dtsouma@ionio.gr}}
\affil[2]{Aristotle University of Thessaloniki, Data Engineering Laboratory, Thessaloniki, Greece\\
  \texttt{tsichlas@csd.auth.gr}, \texttt{akosmato@csd.auth.gr}, \texttt{papadopo@csd.auth.gr}}
  
  \affil[3]{National Technical University of Athens, Computing Systems Lab, Athens, Greece \\ 
  \texttt{katerina@cslab.ece.ntua.gr}}
\authorrunning{S. Sioutas et al.} %mandatory. First: Use abbreviated first/middle names. Second (only in severe cases): Use first author plus 'et. al.'

% mandatory: Please choose ACM 1998 classifications from http://www.acm.org/about/class/ccs98-html . E.g., cite as "F.1.1 Models of Computation". 
% mandatory: Please provide 1-5 keywords
\EventShortTitle{ACM CoRR 2017}
\begin{document}

\maketitle

\begin{abstract}
The skyline of a set $P$ of points ($SKY(P)$) consists of the "best" points with respect to minimization or maximization of the attribute values. A point $p$ dominates another point $q$ if $p$ is as good as $q$ in all dimensions and it is strictly better than $q$ in at least one dimension. In this work, we focus on the static $2$-d space and provide expected performance guarantees for $3$-sided Range Skyline Queries on the Grid, where $N$ is the cardinality of $P$, $B$ the size of a disk block, and $R$ the capacity of main memory. 
We present the MLR-tree (Modified Layered Range-tree), which offers optimal expected cost for finding planar skyline points in a $3$-sided query rectangle,
$q=[a,b] \times(-\infty,d]$, in both RAM and I/O model on the grid 
$[1,M] \times [1,M]$, by single scanning only the points contained in $SKY(P)$. In particular, it supports skyline queries in a $3$-sided range in 
$O\left(t \cdot t_{PAM}\left(N\right)\right)$ time
($O\left((t/B) \cdot t_{PAM}\left(N\right)\right)$ I/Os), where $t$ is the answer size and $t_{PAM}\left(N\right)$ the time required for answering predecessor queries for $d$ in a PAM (Predecessor Access Method) structure, which is a special component of MLR-tree and stores efficiently root-to-leaf paths or sub-paths. By choosing PAM structures with $O(1)$ expected time for predecessor queries under discrete $\mu$-random distributions of the $x$ and $y$ coordinates, MLR-tree supports skyline queries in 
optimal $O(t)$ expected time ($O(t/B)$ expected number of I/Os) with high probability. The space complexity becomes superlinear and can be reduced to linear for many special practical cases. If we choose a PAM structure with $O(1)$ amortized time for batched predecessor queries (under no assumption on distributions of the $x$ and $y$ coordinates), MLR-tree supports batched skyline queries in optimal
$O(t)$ amortized time, however the space becomes exponential.In dynamic case, the update time complexity is affected by a $O(log^{2}N)$ factor. 
\end{abstract}

\section{Introduction} 

In this paper, we study efficient algorithms with non-trivial performance guarantees for {\it skyline} processing on the static plane. 
Let $P$ denote the set of points in the dataset.
Also, let $p_i$ denote the value of the $i$-th coordinate of a point $p$ (in our case $i \in \{1,2\}$). 

\begin{definition}[Dominance]
A point $p \in P$ dominates another point $q \in P$ ($p \prec q$) when $p$ is as good as $q$ in all dimensions and strictly better than $q$ 
in at least one dimension. Formally: $p \prec q$ when $\forall i$, $p_i \leq q_i$ and $\exists j$ such that $p_j < q_j$. 
\end{definition} 

\begin{definition}[Skyline]
The skyline of a set of points $S$ contains the points that are not dominated by any other point. Formally:
$$
SKY(P) = \{p \in P~|~ \nexists q \in P: q \prec p\}
$$
\end{definition}

In the above definitions, we have assumed that small values are preferable. However, this may change according to the concept
and characteristics of the dimensions. For example, if each point represents a purchased item with one dimension being the {\tt price} 
and the other dimension being the {\tt quality}, then the best items should have low price and high quality. 

Skyline queries have attracted the interest of the database community for more than a decade. Although the problem was already known in
Computational Geometry under the name maximal (or minimal) vectors, the necessity to support skyline queries in databases was first
addressed in \cite{BKS01}. Since, they have been used in many applications including multi-criteria decision making, data mining
and visualization, quantitative economics research and environmental surveillance \cite{prefer,miningSkylines,skylinecube}.
Assume that we use the operator {\tt SKYLINE OF} to express skyline queries using SQL. 
Then, a SQL query asking for the skyline of a relation could look like the following two examples: ~\\

\noindent
\indent \indent {\tt SELECT id, name, price, quality} ~\\
\indent \indent {\tt FROM items}~\\
\indent \indent {\tt WHERE price <= 100 AND quality>=3}~\\
\indent \indent {\tt SKYLINE OF price MIN, quality MAX}~\\

\noindent
\indent \indent {\tt SELECT player\_name, Height, Performance} ~\\
\indent \indent {\tt FROM Basketball\_Team}~\\
\indent \indent {\tt WHERE Height IN $[1.90, 2.10]$ AND} ~\\ \indent \indent {\tt Performance IN $[0.8, 1]$} ~\\
\indent \indent {\tt SKYLINE OF Height MAX, Performance MAX}~\\

\begin{figure}[htbp]
	\centering
		\includegraphics[width=0.40\textwidth]{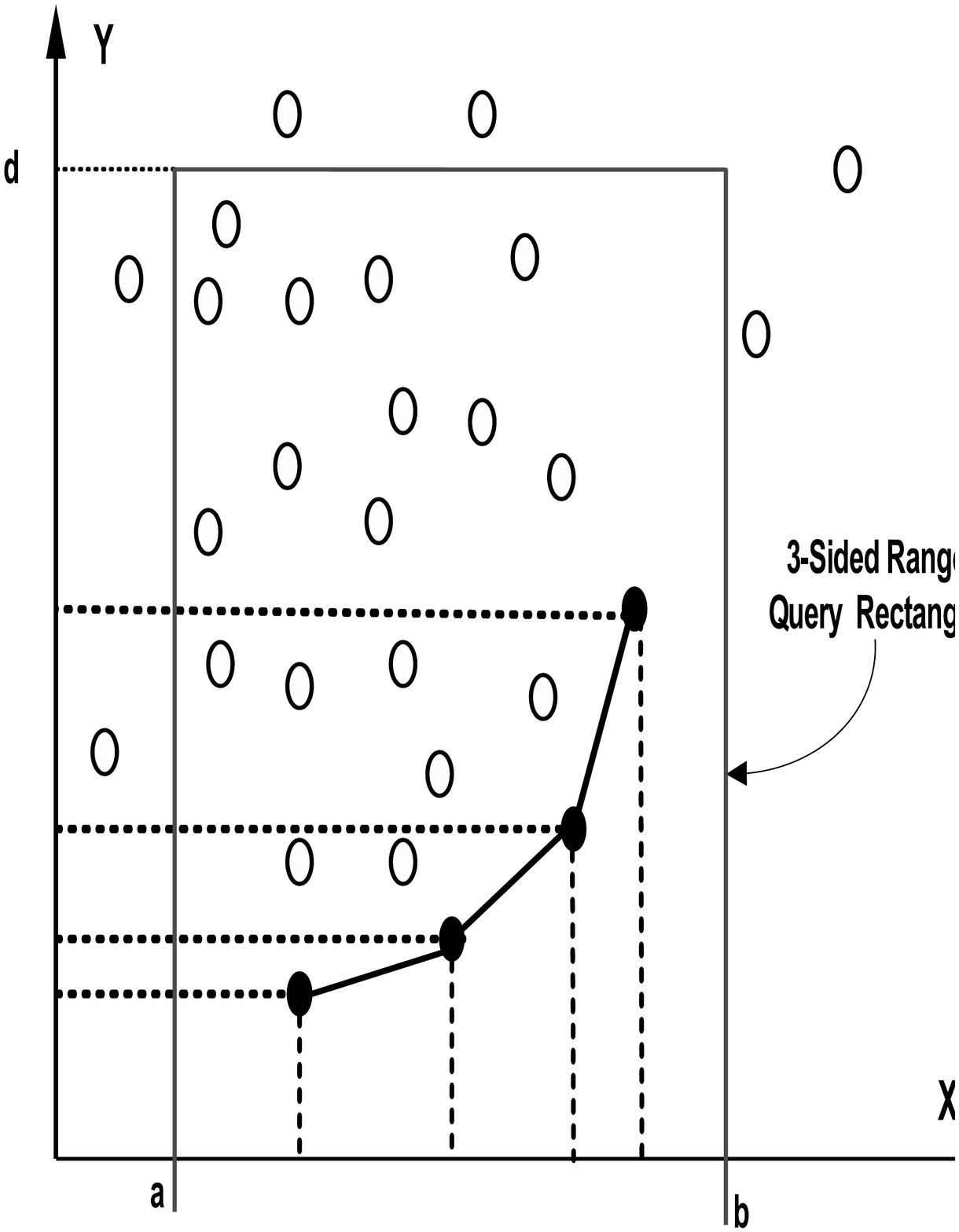}
	\caption{Example of a skyline result in a 3-sided query rectangle}
	\label{fig:maxxminy}
\end{figure}

In first example, price and quality are dependent variables. However, in second example, the two variables, Height of Player and his overall 
Performance respectively, are completely independent. General speaking, in multi-dimensional space, various well known dimensionality
reduction techniques \cite{prefer,miningSkylines,skylinecube} generate spatial vectors with uncorrelated (independent) dimensions.
Thus, the probabilistic study of skyline problem with independent dimensions is of great practical interest.
Observe that, in addition to the skyline preferences the {\tt WHERE} clause contains some additional constraints. For example, 
the user may not be interested in an item that is more expensive than she can afford. Usually, these additional constraints form
a rectangular area referred to as the {\it region of interest}. The answer to the query comprises the skyline of the points 
falling inside the region of interest. Another example is given in Figure \ref{fig:maxxminy}, where MAX(X)-MIN(Y) semantics are being used.
The skyline of the entire dataset is composed of the points $a$, $b$, and $d$, whereas the skyline inside the region of interest
contains the black dot points.

In this paper, we present the MLR (Modified Layered Range) tree-structure providing an optimal expected solution 
for finding planar skyline points in $3$-sided query rectangle in both RAM and I/O model on the grid $[1,M] \times [1,M]$, 
by single scanning not all the sorted points but the points of the answer $SKY(P)$ only. 

The latter means that MLR-tree supports planar skyline queries in a $3$-sided range in 
$O\left(t \cdot t_{PAM}\left(N\right)\right)$ time
($O\left((t/B) \cdot t_{PAM}\left(N\right)\right)$ I/Os), where $t$ is the answer size and $t_{PAM}\left(N\right)$ the time required for answering predecessor queries for $d$ in a PAM (Predecessor Access Method) structure, which is a special component of MLR-tree and stores efficiently root-to-leaf paths or sub-paths. By choosing PAM structures with $O(1)$ expected time for predecessor queries under descrete $\mu$-random distributions of the $x$ and $y$ coordinates, MLR-tree supports skyline queries in 
optimal $O(t)$ expected time ($O(t/B)$ expected number of I/Os) with high probability. In addition to the general case, where the $x$ and $y$ coordinates are drawn from a $\mu$-random distribution, we examine two more special cases with practical interest:
(a) The inserted points have their $y$-coordinates drawn from a class of $(f_1, f_2)-smooth$ distributions, whereas the $x$-coordinates 
are arbitrarily distributed. In this case the space becomes linear, also the query time is marginally affected by a very small sublogarithmic factor.
(b) The $x$-coordinates are arbitrarily distributed and the $y$-coordinates are continuously drawn from a more restricted class of smooth distributions. Similarly, the space is reduced to linear, also the query time remains unaffected (optimal).
The practical interest of these special cases stems from the fact that {\em any} probability distribution is $(f_{1},\Theta(n))$-smooth, 
for a suitable choice of the parameters, as we will describe later. Finally, if we choose a PAM structure with $O(1)$ amortized time for batched predecessor queries (under no assumption on distributions of the $x$ and $y$ coordinates), MLR-tree supports batched skyline queries in optimal
$O(t)$ amortized time, however the space becomes exponential.In dynamic case, the update time complexity is affected by a $O(log^{2}N)$ factor. 

The proposed data structure borrows ideas from the Modified Priority Search Tree presented in \cite{KPSTT10} 
that supports simple 3-sided range reporting queries. However, the modifications to support skyline queries are 
novel and non-trivial. The same problem (dynamic I/O-efficient range skyline reporting) but with worst case guarantees 
(logarithmic query I/Os, logarithmic amortized update I/Os and linear space) has been presented in \cite{RTTTY13}.

The rest of the work is organized as follows. Related work in the area and a brief discussion of our contributions are given in Section\ref{sec:rw}, whereas some fundamental concepts are presented in Section \ref{sec:fundamental}. A detailed description and analysis of our contributions are given in Section \ref{sec:mlr}. Finally, in Section \ref{sec:conclusions} we conclude the work and discuss future research briefly.

%The skyline is a subset of the convex hull \cite{PS85} set, which is a tougher problem. The best known algorithm to compute
%the convex hull in limited main memory has a complexity of $O(n^{\left\lfloor d/2 \right\rfloor +1})$ for $d>3$ and $O(nlogn)$ for $d=2,3$. 
%Optimal solutions for planar 3-sided range maxima queries in RAM and Comparison (Pointer Machine) model were presented in \cite{BT11}. 
%In I/O model,the best previous solution for multi-dimensional skyline (maxima) queries presented in \cite{ST11} 
%and the best practical (implementable) solution presented in \cite{PTFS03} and \cite{PTFS05} by Papadias et al.. 

%In distributed (P2P) and parallel model the best solutions have been presented in \cite{VP10} and \cite{HJS11} respectively. 

\section{Related Work and Contributions}
\label{sec:rw}

In this section, we describe related research in the area, focusing on the best available results. In addition, we present our contributions.\\

\textbf{Results for RAM and PM models:} 
The best previous solution presented in \cite{BT11} and supports maxima (skyline) queries in optimal $O(\log N / \log log N+t)$ worst case time and updates in 
$O(\log N / \log log N)$ worst case time 
consuming linear space in the RAM model of computation with word size $w$, where the coordinates of the points are integers in the range $U={0, \cdots , 2^{w-1}}$. 

In the Pointer Machine (PM) or Comparison model (comparison is the only allowed computation on the coordinates of the points) 
the solution in \cite{BT11} requires optimal $O(\log N + t)$ worst case query time and $O(\log N)$ worst case update time. 
The data structure of \cite{BT11} also supports the more general query of reporting the maximal points among the points that lie in a given 3-sided orthogonal range unbounded from above in the same complexity. It can also support
4-sided queries in $O(\log^2 N+t)$ worst case time, and $O(\log^2 N)$ worst case update time, using $O(N \log N)$ space, where $t$ is the size of the output.
\\
\textbf{Results for the I/O model:}
The best previous solution has been presented in \cite{ST11}. In the external-memory
model, the 2-d version of the problem is known to be solvable in $O((\frac{N}{B}) \log_{R/B}\frac{N}{B} + \frac{N}{B} )$ I/Os 
and $O(N/B)$ (i.e., linear) space, where $N$ is the cardinality of $P$, $B$ the size of a disk block, and $R$ the capacity of main memory. 

%\begin{figure}[htbp]
%%\includegraphics[width=0.35\textwidth]{2D-Sky-Sorting.eps}
%\epsfig{file=2D-Sky-Sorting.eps,width=5cm} %0.35\columnwidth}
%\caption{Illustration of algorithms by Kung et al.\protect \cite{KLP75}}
%\label{fig:2D-Sky-Sorting}
%\end{figure}

In particular, the skyline $SKY(P)$ of a set $P$ of 2-d points can be
extracted by a single scan, provided that the points of $P$ have
been sorted in ascending order of their $x$-coordinates. For example,
consider any point $p \in P$; and let $P'$ be the set of points of $P$ 
that rank before $p$ in the sorted
order. Apparently, $p$ cannot be dominated by any point that
ranks after $p$, because $p$ has a smaller $x$-coordinate than any
of those points. On the other hand, $p$ is dominated by some
point in $P'$ if and only if the $y$-coordinate of $p$ is greater
than $y_{min}$, where $y_{min}$ is the smallest $y$-coordinate of all the
points in $P'$.

\begin{table*}[!t]

Table 1: New bounds for dynamic $3$-sided planar skyline on the grid $[1,M] \times [1,M]$ in RAM model of computation.\\
{
	\centering
		\begin{tabular}{|l||l||l|} \hline
 Data Distributions & Space & Query Time \\ \hline \hline
   $x$: \emph{$\mu$-random} & $O(N^{1+\delta}+M)$ & $O(t)$ expected  \\ 
   $y$: \emph{$\mu$-random}         &                         &    \\ \hline
   $x$: \emph{$\mu$-random}  & $O(N+M)$ & $O(t loglogN)$ expected \\
   $y$: $(f_1,f_2)-smooth$   &              &                        \\ \hline
   $x$: \emph{$\mu$-random}  & $O(N+M)$ & $O(t)$ expected \\
   $y$: $(\frac{N}{g(N)}, ln^{O(1)}N)-smooth$ &    &     \\ \hline
   $x$: \emph{arbitrary}  & $O(N^{p(M)})$ & $O(t)$ amortized \\
   $y$: \emph{arbitrary}   &  for any $p(M)=O(loglogM)$  & for $p(M)$ batched skyline queries     \\ \hline
		\end{tabular}
}
\end{table*}

\begin{table*}[!t]
Table 2: New bounds for dynamic $3$-sided planar skyline on the grid $[1,M] \times [1,M]$ in I/O model of computation.\\
{
	\centering
		\begin{tabular}{|l||l||l|} \hline
 Data Distributions & Space & Query Time \\ \hline \hline
   $x$: \emph{$\mu$-random}  & $O((N+M)/B)$ & $O((t/B)log_{B}logN)$ expected \\
   $y$: $(f_1,f_2)-smooth$   &              &                        \\ \hline
   $x$: \emph{$\mu$-random}  & $O((N+M)/B)$ & $O(t/B)$ expected \\
   $y$: $(\frac{N}{g(N)}, ln^{O(1)}N)-smooth$ &    &     \\ \hline
		\end{tabular}
}
\end{table*}

To populate $SKY(P)$, it suffices to read $P$ in its sorted order, and
at any time, keep the smallest $y$-coordinate $y_{min}$ of all the
points already seen. The next point $p$ scanned is added to
$SKY(P)$ if its $y$-coordinate is below $y_{min}$, in which case $y_{min}$
is updated accordingly. In the I/O model, this algorithm
performs $O((N/B) log_{R/B}(N/B))$ I/Os, which is the time
complexity of sorting $N$ elements in external memory.

For fixed $d\geq3$, the solution presented in \cite{ST11} 
requires $O((\frac{N}{B}) \log_{R/B}^{d-2}\frac{N}{B} + \frac{N}{B} )$ I/Os. Previously,
the best solution was adapted from an in-memory algorithm,
and requires $O((\frac{N}{B}) \log_{2}^{d-2}\frac{N}{B} + \frac{N}{B} )$ I/Os.\\

%\spara{Practical solutions}.
%The best practical solutions for skyline queries are reported in \cite{PTFS05}. 
%The BBS algorithm works over an R-tree index and supports incremental computation of the skyline result.
%However, no theoretical guarantee is given with respect to the number of I/Os that are required. 

\textbf{Our Contributions:} In this work, we provide novel algorithmic techniques with non-trivial performance guarantees, to process planar skyline queries inside a region of interest. Evidently, in its static case (no insertions and deletions),
the problem can be solved by reusing existing techniques that return the skyline of the entire dataset and keeping only
the points that fall inside the region of interest. This approach leads to suboptimal solutions, since the 
processing cost does not depend on the size of the region and the number of points that fall inside: for every query,
the whole dataset must be scanned. In addition, most of the proposed algorithms are not equipped to handle insertions and deletions of points.

An exception to this behavior is the BBS \cite{PTFS05} algorithm, which in most cases returns the skyline without scanning the entire R-tree index and in addition, it supports skyline computation inside a region of interest
and can handle insertions and deletions. However, BBS does not offer any theoretical performance guarantee.

In this paper, we propose the MLR (Modified Layered Range) tree-structure 
providing an optimal expected solution for finding planar skyline points in a given $3$-sided query rectangle 
in both RAM and I/O model on the grid $[1,M] \times [1,M]$, by single scanning only the points contained in $SKY(P)$. 
The latter means that the MLR-tree supports planar skyline queries in $O(t/B)$ expected number of I/Os ($O(t)$ in RAM), 
where $t$ the answer cardinality, consuming also super-linear space (general case), which becomes linear under specific data distributions. Also, MLR-tree supports batched skyline queries in optimal $O(t)$ amortized time, however the space becomes exponential. In dynamic case, the update time complexity is affected by a $O(log^{2}N)$ factor.Our results are summarized in Tables 1 and 2.
 %\ref{contrib}.

\section{Fundamental Concepts}
\label{sec:fundamental}

For main memory solutions we consider the RAM model of computation.
We denote by $N$ the number of elements that reside in the data structures and by $t$ the size of the query. 

For the external memory solutions we consider the I/O model of computation \cite{V01}. This means that the input resides in the external memory in a blocked fashion. Whenever a computation needs to be performed to an element, the block of size $B$ that contains that element is transferred into main memory, which can hold at most $R$ elements. Every computation that is performed in main memory is free, since the block transfer is orders of magnitude more time consuming. Unneeded blocks that reside in main memory are evicted by a LRU replacement algorithm. Naturally, the number of block transfers (\emph{I/O operation}) consists the metric of the I/O model.

Furthermore, in the dynamic case we will consider that the points to be inserted are drawn by an unknown descrete distribution. Also, the asymptotic bounds are given with respect to the current size of the data structure. Finally, deletions
of the elements of the data structures are assumed to be uniformly random. That is, every element present in the data structure is
equally likely to be deleted \cite{K77}.

\subsection{Probability Distributions}
In this section, we overview the probabilistic distributions that will be used in the remainder of the paper. 
We will consider that the $x$ and $y$-coordinates are distinct elements of these distributions and will choose
the appropriate distribution according to the assumptions of our constructions.

A probability distribution is \emph{$\mu$-random} if the elements are drawn randomly with respect
to a density function denoted by $\mu$. For this paper, we assume that $\mu$ is unknown.

Informally, a distribution defined over an interval $I$ is
\textit{smooth} if the probability density over any subinterval of $I$
does not exceed a specific bound, however small this subinterval is
(i.e., the distribution does not contain sharp peaks). 

Given two functions $f_{1}$ and $f_{2}$, then $\forall x \in U, \mu(x)$ is {\em $(f_{1},f_{2})$-smooth} if there exists a constant $\beta$, such that for all $c_{1},  c_{2},  c_{3}\in U: c_{1} < c_{2} <c_{3}$, and for all naturals $ \nu \leq n$, for a random key $x \in U$ it holds that:
\begin{eqnarray}
\label{Def:smooth-discrete}
\Pr\left[c_{2}-\left\lfloor \frac{c_{3}-c_{1}}{f_{1}(\nu)}\right\rfloor \leq x \leq c_{2} | c_1 \leq x \leq c_3\right]
=\sum_{c_{2}- \left\lfloor\frac{c_{3}-c_{1}}{f_{1}(\nu)}\right\rfloor}^{c_{2}}
{\mu_{[c_{1},~ c_{3}]}(x)}
   \leq \frac{\beta f_{2}(\nu)}{\nu}~
\end{eqnarray}
where $\mu_{[c_{1},c_{3}]}(x)= 0$ for $x < c_1$ or $x > c_3$, and $\mu_{[c_{1},~c_{3}]}(x)=\mu(x)/p$ for $x \in \{c_1, \ldots, c_3\}$ where
$p=\sum_{c_1}^{c_3} \mu(x)$ and $p>0$.

The above imply that no key can get a point mass, i.e. a value with nonzero\footnote{In the sense that it is bounded below by a positive constant.} probability. More accurately, if we initially consider the whole universe of keys with $|U|= n^c, c>1$, and $\forall \nu \leq n$, we equally split it into $f_1 (\nu)= \nu^{\alpha}, \alpha < 1$, many equal consecutive subsets of keys, then (\ref{Def:smooth-discrete}) implies that each subset (containing $\geq \frac{|U|}{f_1 (\nu)}= \frac{n^c}{\nu^{\alpha}}= \omega(1)$ consecutive keys) gets probability mass $\leq \frac{f_2(\nu)}{\nu}= \frac{\nu^{\delta}}{\nu}, \forall \nu \leq n$, which is $o(1)$ as $n \rightarrow \infty$, when $f_2(n)= n^{\delta}, \forall \delta \in (0, 1)$. Hence, as $n \rightarrow \infty$, each key in $U$ has $o(1)$ probability mass. Once more, we can describe (\ref{Def:smooth-discrete}) by rephrasing the intuitive description of $(f_1, f_2)$-smooth distribution as:

``{\it among a number (measured by $f_1(n)= n^{\alpha}, \alpha < 1$) of consecutive subsets, each containing consecutive keys from $U$,  no subset containing consecutive keys from $U$ should be too dense (measured by $f_2(n)= n^{\delta}, \delta < 1$) compared to the others}''.

The class of $(f_{1},f_{2})$-smooth distributions (for appropriate choices of
$f_1$ and $f_2$) is a superset of both regular and uniform classes of
distributions, as well as of several non-uniform classes
\cite{AM93,KMSTTZ03}. Actually, {\em any} probability distribution is
$(f_{1},\Theta(N))$-smooth, for a suitable choice of $\beta$.

The \emph{grid distribution} assumes that the elements are integers that belong to a specific range $[1,M]$.

\subsection{Preliminary Access Methods} \label{ssec:DS}

In this section, we describe the data structures that we utilize in order to achieve the desired complexities.\\

\textbf{Half-Range Minimum/Maximum Queries:} The {\em half-Range Maximum Query} (h-RMQ) problem asks to preprocess an array $A$ of size $N$ such that,
given an index range $[1,r]$ where $1\leq r \leq N$, we are asked to report the position of the maximum element in this range on $A$. Notice that we do not want to change the order of the elements in $A$, in which case the problem would be trivial.
This is a restricted version of the general RMQ problem, in which the range is $[r,r']$, where $1\leq r\leq r'\leq N$. In~\cite{HT84} the RMQ problem is solved in $O(1)$ time using $O(N)$ space and $O(N)$ preprocessing time. The currently most space efficient solution that supports queries in $O(1)$ time appears in~\cite{FH07}. We could use these solutions for our h-RMQ problem, but in our case the problem can be solved much simpler by maintaining an additional array $A_{max}$ of maximum elements for each index of the initial array.

\textbf{The Lazy B-tree:} The Lazy B-tree of \cite{KMMSTTZ05} is a simple but non-trivial externalization of the techniques introduced in \cite{R92}. 
The first level consists of an ordinary B-tree, 
whereas the second one consists of buckets of size $O(\log^2 N)$, where $N$ is approximately equal to the number of elements stored 
in the access method. 
Each bucket consists of two list layers, $L$ and $L_{i}$ respectively, where $1\leq i \leq O(\log N)$, each of which has $O(\log N)$ size. 
The technical details concerning both the maintenance of criticalities and the representation of buckets, can be found in \cite{KMMSTTZ05}. 
The following theorem provides the complexities of the Lazy B-tree:
\\

\textbf{Theorem1:} The Lazy B-Tree supports the search operation in $O(\log_B N)$ worst-case block transfers and update operations in $O(1)$ worst-case block transfers, provided that the update position is given.
\\

%\begin{theorem}[LBT]
%The Lazy B-Tree supports the search operation in $O(\log_B N)$ worst-case block %transfers and update operations in $O(1)$ worst-case block transfers, provided that %the update position is given.
%\end{theorem}

\textbf{Interpolation Search Trees:} In \cite{KMSTTZ06}, a dynamic data structure based on interpolation search (IS-Tree) was presented, which requires linear space and can be updated in $O(1)$ time w.c. Furthermore, the elements can be searched
in $O(\log \log N)$ time expected w.h.p., given that they are drawn from a $(N^{\alpha}, N^{\beta})$-smooth distribution,
for any arbitrary constants $0< \alpha , \beta < 1$. The externalization of this data structure, called interpolation search B-tree (ISB-tree), was introduced in \cite{KMMSTTZ05}. It supports update operations in $O(1)$
worst-case I/Os provided that the update position is given and search operations in $O(\log_B \log N)$ I/Os expected w.h.p. The expected search bound holds w.h.p. if the elements are drawn by a $(N/(\log \log N)^{1+\epsilon}, N^{1-\delta})$-smooth distribution, where $\epsilon > 0$ and $\delta = 1 -\frac{1}{B}$ are constants. If the elements are drawn by the more restricted $(\frac{N}{g}, ln^{O(1)}N)-smooth$ densities the expected number of I/Os for the search operation becomes $O(1)$ with high probability ($g$ is an arbitrarily chosen constant). The worst case search bound is $O(\log_{B}N)$ block transfers.
\\
\textbf{Random Input:} The Data Structure presented in \cite{BKS11}, the Random Search Array (RSA), alleviates all lower bounds for the dynamic predecessor search problem, 
by proving constant time with high probability (w.h.p.), as $N$ grows large, thus, improving over all approaches presented in \cite{MT93,AM93,KMSTTZ06}. 
The fine details of this dynamic data structure exhibit that achieves constant predecessor time w.h.p., working with only $O(N)$ short memory words of length $w$-bits, meaning that $0\leq N \leq 2^{w}-1$ or $w\geq logN$. For $w$ equals to
exactly $logN$-bits and for $0<\delta<1$, RSA consumes super-linear space $O(N^{1+\delta})$. The tuning of positive constant $\delta$ for practical purposes was not studied in this paper. 
%Obviously, the RSA in external memory (block fashion) requires $O(\frac{N^{1+\delta}}{B})$ disk-blocks.
\\
\textbf{\textbf{B}atched \textbf{P}redecessor \textbf{Q}ueries:} The Data Structure presented in \cite{MKYN05}, answers batched predecessor queries in $O(1)$ amortized time. In particular, it supports $O(\sqrt{logN})$ queries in $O(1)$ time per query and requires $O(N^{\epsilon\sqrt{logN}})$ space for any $\epsilon > 0$, where $M$ is the size of the universe. It also can answer $O(\log\log M)$ predecessor queries in $O(1)$ time per query and requires $O(N^{\epsilon loglogM})$ space for any $\epsilon > 0$. The method of solution relies on a certain way of searching for predecessors of all elements of the query in parallel.

In a general case, the solution in \cite{MKYN05} presents a data structure that supports $p(N)$ queries in $O(\frac{\sqrt{logN}}{p(N)})$ time per query and requires $O(N^{p(N)})$ space for any  $p(N)=\sqrt{logN}$, as well as a data structure that supports $p(M)$ queries in $O(\frac{loglogM}{p(M)})$ time per query and requires $O(N^{p(M)})$ space for any  $p(M)=loglogM$.

\section{The MLR-TREE} \label{sec:mlr}

In the following, we describe in detail the indexing scheme, which is termed the {\it Modified Layered Range Tree} (MLR-tree). The description of the MLR-tree is considered in the MAX-X, MIN-Y case. The other three cases can be handled in a similar way.

\subsection{The Main Memory Static Non-Linear-Space MLR-tree} \label{ssec:nonlinearmlr}

The Static Non-Linear MLR-tree (see Figure ~\ref{fig:mlrtreeram}) is a static data structure that stores points on the 2-d grid. 
It is stored as an array $(A)$ in memory, yet it can be visualized as a complete binary tree. 
The static data structure is an augmented binary search tree $T$ on the set of points $S$ that resembles a range tree. $T$ stores all points in its leaves with respect to their $x$-coordinate in increasing order.  Let $H$ be the height of tree $T$. We denote by $T_{v}$ the subtree of $T$ with root the internal node $v$. 

Let $P_{\ell}$ be the root-to-leaf path for leaf $\ell$ of $T$. We denote by $P^{\tau}_{\ell}$ the subpath of $P_{\ell}$ consisting of nodes with depth $\geq {\tau}$. Similarly, $P^{\tau}_{\ell,left}$ ($P^{\tau}_{\ell,right}$) denotes the set of nodes that are left (right) children of nodes of $P^{\tau}_{\ell}$ and do not belong to $P^{\tau}_{\ell}$. Let $q=(q_x,q_y)$ be the point stored in leaf $\ell$ of the tree  where $q_x$ is its $x$-coordinate and $q_y$ is its $y$-coordinate. $P_{q}$ denotes the search path for $q_x$, i.e., it is the path from the root to $\ell$ and it is equal to $P_{\ell}$. The binary tree is augmented as follows:

%\begin{figure}[htbp]
\begin{figure}[!t]
	\centering		
		\includegraphics[width=1.0\textwidth]{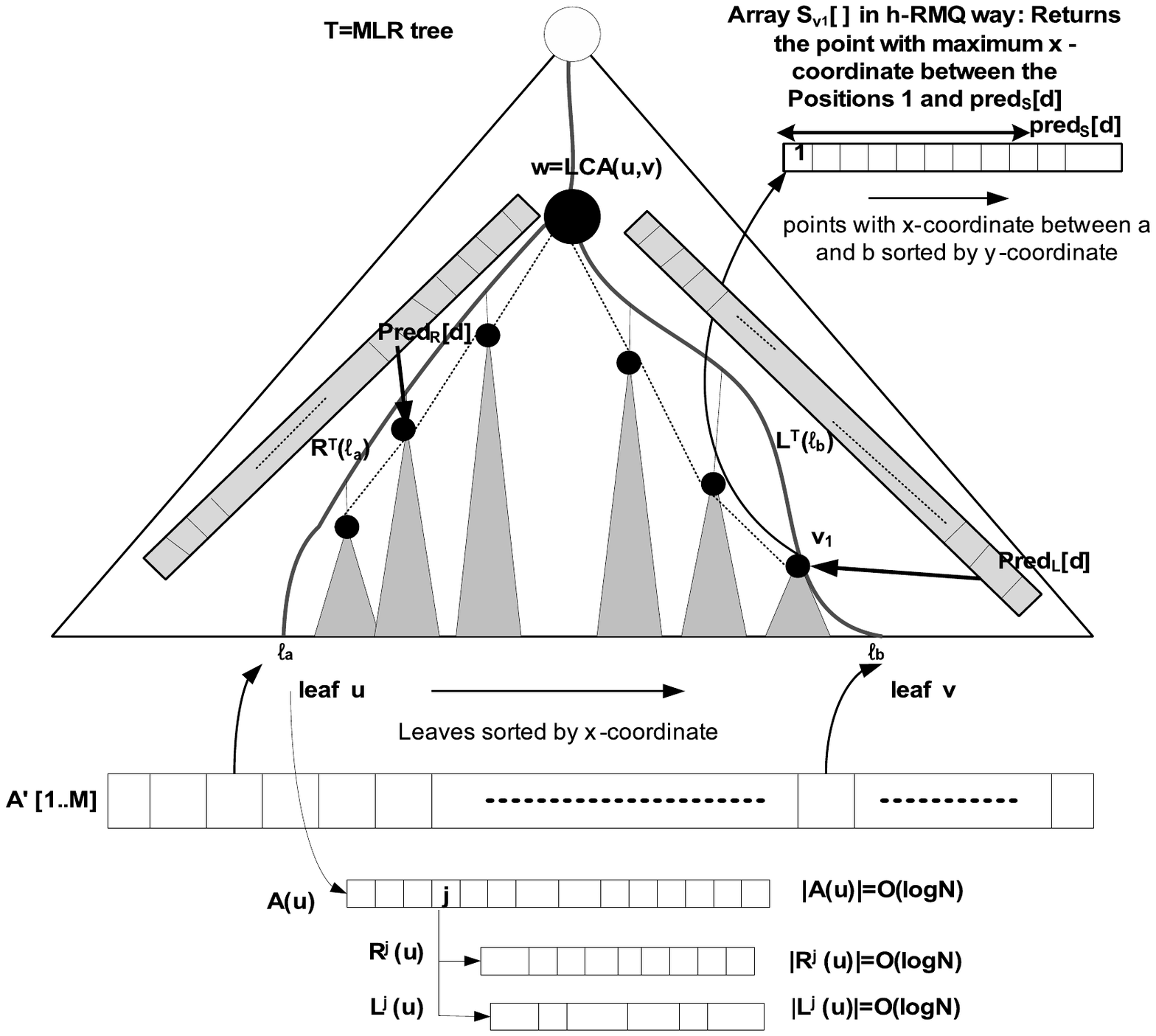}
		\vspace{-3cm}
	\caption{The static non-linear-space MLR-tree in main memory.}
	\label{fig:mlrtreeram}
\end{figure}

\begin{itemize}

\item Each internal node $v$ stores a point $q_v$, which is the point with the minimum $y$-coordinate among all points in its subtree $T_v$.

\item Each internal node $v$ is equipped with a secondary data structure $S_v$, which stores all points in $T_v$ with respect to $y$-coordinate in increasing order. $S_v$ is implemented with a Predecessor Access Method (PAM) as well as an h-RMQ structure (see ~\ref{ssec:DS}).

\item Each leaf $\ell$ stores arrays $L^{\tau}_{\ell}$ and $R^{\tau}_{\ell}$, where $0\leq {\tau} \leq H-1$, corresponding to sets $P^{\tau}_{\ell, left}$ and $P^{\tau}_{\ell,right}$ respectively. In particular, these arrays contain the points $q_v$ for each node $v$ in the corresponding sets. These arrays are sorted with respect to their $y$-coordinate and are implemented with a PAM. In addition, they are also implemented as h-RMQ structures. 
\end{itemize}

We use an array $A'$ of size $M$, which stores pointers to the leaves of $T$. In particular, $A'\left[q_x\right]$ contains a pointer to the leaf of $T$ with maximum $x$-coordinate smaller or equal to $q_x$ (this is $q_x$'s predecessor). In this way, we can determine in $O(1)$ time the leaf of a search path for a particular point in $T$. Finally, tree $T$ is preprocessed in order to support Lowest Common Ancestor queries in $O(1)$ time. Since $T$ is static, one can use the methods of~\cite{G94,HT84} to find the LCA (as well as its depth) of two leaves in $O(1)$ time by attaching to each node of $T$ a simple label.

Having concluded with the description of the data structure, we move to the skyline query. Assume we want to compute the skyline in the query range $q=[a,b] \times [-\infty,d]$. The procedure to compute the points on the skyline is the following:

\begin{enumerate}
\item We use the array $A'$ to find the two leaves $\ell_a$ and $\ell_b$ of $T$ for the search paths $P_{a}$ and $P_{b}$ respectively. Let $w$ be the LCA of leaves $\ell_a$ and $\ell_b$ and let ${\tau}$ be its depth. \label{step:1}

\item The predecessor of $d$ is located in $R^{\tau}(\ell_a)$ and $L^{\tau}(\ell_b)$ and let these predecessors be at positions $pred_L[d]$ and $pred_R[d]$ respectively. In addition, let $\nu_1$ be the node that has the following property: the $y$-coordinate of point $q_{\nu_1}$ belongs in the range $[-\infty,d]$ and it has the largest $x$-coordinate (the $x$-coordinate of $q_{\nu_1}$ falls in the $[a,b]$ range because of step~\ref{step:1}) among all nodes in $P^{\tau}_{\ell_a,right}$ and $P^{\tau}_{\ell_b,left}$. This means that node $\nu_1$ is the rightmost node that has a point with $y$-coordinate within the range $(-\infty,d]$. \label{step:2}

\item By executing an h-RMQ in $L^{\tau}_{\ell_b}$ and $R^{\tau}_{\ell_a}$ arrays for the range $[1,pred_L[d]]$ and $[1,pred_L[d]]$ node $\nu_1$ is located. The subtree $T_{\nu_1}$ stores the point (which surely exists) with the maximum $x$-coordinate among all points in the query range $[a,b]\times (-\infty,d]$. By executing a predecessor query for $d$ in $S_{\nu_1}$ returning the result $pred_S[d]$, and then making an h-RMQ in $S_{\nu_1}$ for the range $[1,pred_S[d]]$, we find and report the required point with the maximum $x$-coordinate $z=(z_x,z_y)$ that belongs to the skyline (recall that we use MAX-X and MIN-Y semantics). \label{step:3}

\item The query range now becomes $q=[a, z_x]\times (-\infty, z_y]$. 

\item We repeat the previous steps until $S \cap q=\oslash$.

\end{enumerate}

Before moving to the analysis of the data structure we need to prove its correctness with respect to the skyline range query. 

\textbf{Theorem 2:}The skyline range query correctly returns the skyline within the range $[a,b]\times (-\infty,d]$.

%\begin{lemma} \label{lem:correctness}
%The skyline range query correctly returns the skyline within the range $[a,b]\times (-%\infty,d]$. 
%\end{lemma}

\begin{proof}
We prove by induction that the query algorithm returns the point in the skyline in decreasing order with respect to $x$-coordinate. The first time that the algorithm is executed, the point on the skyline with the largest $x$-coordinate is returned. To prove this statement assume that some other point with largest $x$-coordinate is returned. This means that this point should be in a subtree rooted not at $\nu_1$ but at a different node. However, this is impossible since $\nu_1$ is the rightmost subtree whose point with minimum $y$-coordinate is in the range $(-\infty,d]$. For the same reason, $\nu_1$ will be located correctly in the h-RMQ. As a result, the point on the skyline with the largest $x$-coordinate is correctly located and reported first. Assume that some points of the skyline have already been reported. We have to show the following: a) the points considered in the current loop are those in $P^{\tau}_{\ell_a,right}$ and $P^{\tau}_{\ell_b,left}$ whose $y$-coordinate is in the range $(-\infty,d]$ are the ones we must consider and only these and b) the reported point on the skyline has the largest $x$-coordinate among all points in the new query range. For the second part a similar discussion as in the previous paragraph applies. For the first part, it is enough to note that all points that are dominated by the reported skyline points are not considered since the query range has changed.
\end{proof}

Let $S_{PAM}(N)$ be the required space for $N$ elements for the Predecessor Access Method (PAM) and let $t_{PAM}(N)$ be the time complexity for a predecessor query. Finally, let $C_{PAM}(N)$ be the time complexity for the construction of the PAM on $N$ elements. Building tree $T$ is performed in a bottom-up manner. In particular, tree $T$ as well as the respective points within the internal nodes can be built in $O(N\log{N})$ time, since we have to sort the points with respect to the $x$-coordinate. Arrays $S_v$, for all internal nodes $v$, are constructed in a bottom-up manner by merging the two already sorted with respect to $y$-coordinate arrays of the children into one array in their father in linear (to their size) time. Note that the elements are copied and the arrays of the children are not destroyed. Then, we construct the h-RMQ structure in linear time as well as the PAM in $O(C_{PAM}(|T_v|))$ time for node $v$. This can be carried out in $O(N\log{N})$ time since $N$ elements are processed at each level of the tree $T$ as well as in $O(C_{PAM}(N)\log{N})$ time for the PAM of each $S_v$ structure. Finally, sequences $L^{\tau}_{\ell}$ and $R^{\tau}_{\ell}$, where $0 \leq {\tau} \leq H-1$, for all leaves $\ell$, can be constructed one by one in $O(N\log^2{N})$. This is because, for each leaf $\ell$ among the $N$ leaves in total, we construct $\log{N}$ such sequences each of which has size $O(\log{N})$. Each such sequence must be structured with a Predecessor Access Method (PAM) as well as as an h-RMQ structure. In this particular case we choose to use $q^*$-heaps~\cite{WILLARD2000} as a PAM due to the small size of the sets and their linear time construction. The total time to construct the data structure on $N$ elements is $O((C_{PAM}(N)+N\log{N})\log{N})$. 

Recall that for each point of the SKY(P) set, we execute in total three predecessor queries (two of them in Step~\ref{step:2} and one in Step~\ref{step:3}). Since all other steps can be carried out in $O(1)$ time, the total time complexity of the query algorithm is $O(t\times t_{PAM}(\log{N}))$. The space complexity of the MLR-tree is dominated by the space used for implementing the $L^{\tau}_{\ell}$, $R^{\tau}_{\ell}$ and $S_v$ sets as well as by the array $A'$, which is $O((S_{PAM}(N)+N\log{N})\log{N}+M)$ as implied by the discussion in the previous paragraph.

\subsection{The Main Memory Static Linear-Space MLR-tree}\label{ssec:staticlinearMLR}
We can reduce the space of the data structure described in~\ref{ssec:nonlinearmlr} by using pruning techniques as in~\cite{FMNT87,O88}. However, pruning alone does not reduce the space to linear. We can get a better space complexity by recursive pruning until reaching a tree of constant size, but it will still be superlinear by an iterated logarithm\footnote{The iterated logarithm, written as $\log^*{N}$, is equal to the number of times the logarithm must be iteratively applied on $N$ before the result is $\leq 1$ for the first time.} (aggravating by a similar multiplicative term the time complexity of the query). To get an optimal space bound we use a combination of pruning and table lookup, which ends the recursion prematurely. 

%DESCRIPTION OF STRUCTURE
The pruning method is as follows: consider the nodes of $T$ with height $2 \log{\log{N}}$. These nodes are roots of subtrees of $T$ of size $O(\log^2{N})$ and there are $\Theta\left(\frac{N}{\log^2{N}}\right)$ such nodes. Let $T_1$ be the tree whose leaves are these nodes and let $T^i_2$ be the subtrees of these nodes for $1 \leq i \leq \Theta\left(\frac{N}{\log^2{N}}\right)$. We call $T_1$ the first layer of the structure and the subtrees $T^i_2$ the second layer.

\begin{figure}[!t]
	\centering		
		\includegraphics[width=1.0\textwidth]{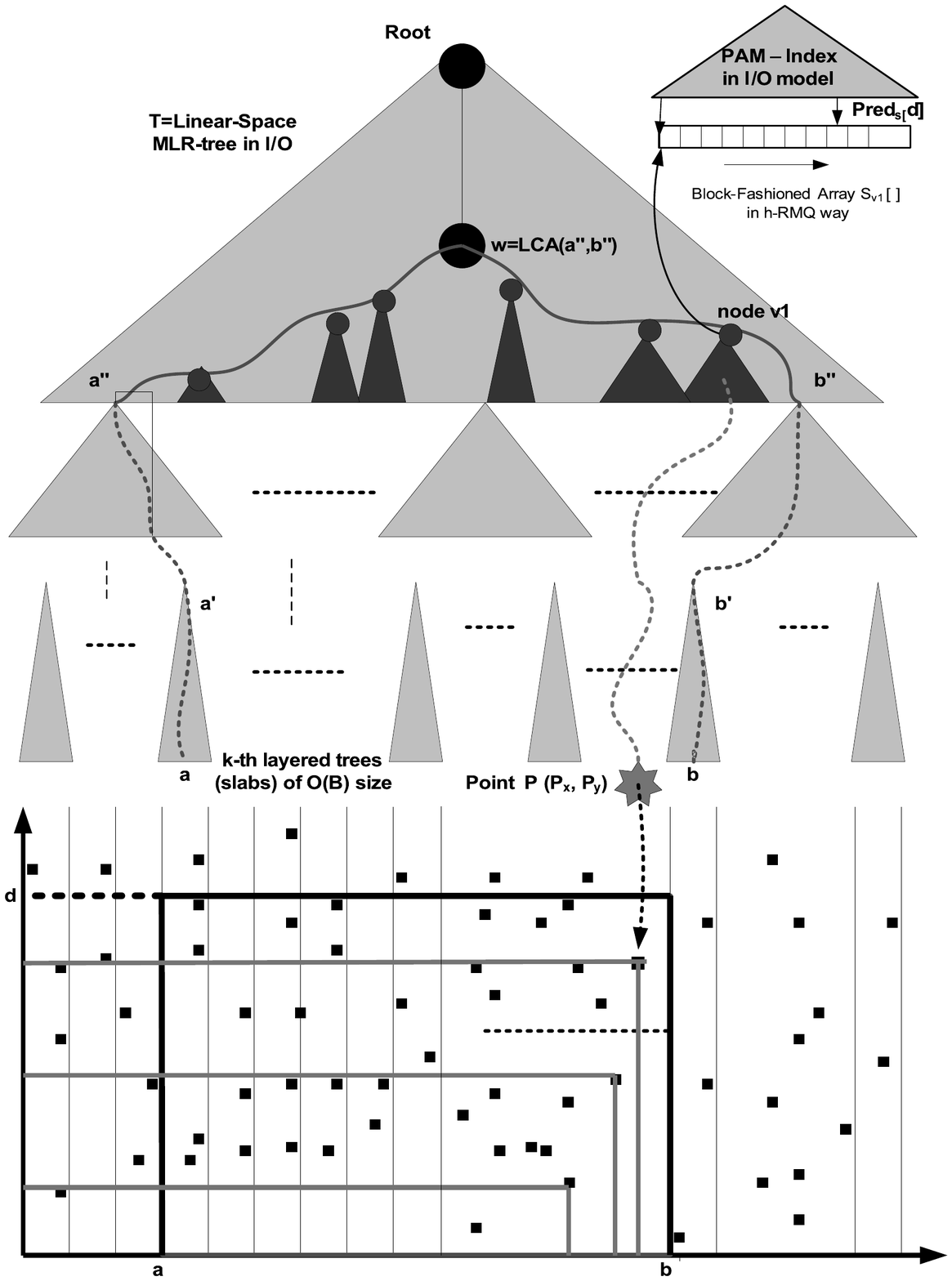}
		\vspace{-3cm}
	\caption{The static linear-space MLR-tree.}
	\label{fig:linearmlr}
\end{figure}

$T_1$ and each subtree $T^i_2$ is implemented as an MLR-tree. The representative of each tree $T^i_2$ is the point with the minimum $y$-coordinate among all points in $T^i_2$. The leaves of $T_1$ contain only the representatives of the respective trees $T^i_2$. Each tree $T^i_2$ is further pruned at height $2\log{\log{\log{N}}}$ resulting in trees $T^j_3$ with $\Theta(\log^2{\log{N}})$ elements. Once again, $T^i_2$ contains the representatives of the third layer trees in a similar way as before. Each tree $T^j_3$ is structured as a table which stores all possible precomputed solutions. In particular, each $T^j_3$ is structured by using a PAM with respect to $x$-coordinate as well as with respect to $y$-coordinate (two different structures in total). In this way, we can extract the position of the predecessor in $T^j_3$ with respect to $x$ and $y$ coordinates. What is needed to be computed for $T^j_3$ is the point with the maximum $x$-coordinate that lies within a $3$-sided range region. To accomplish this, we use precomputation and tabulation for all possible results. 

For the sake of generality, assume that the size of $T^j_3$ is $k$. Let the points in $T^j_3$ be $q_1,q_2,\ldots,q_k$ sorted by $x$-coordinate. Let their rank according to $y$-coordinate be given by the function $\alpha(i), 1\leq i \leq k$. Apparently, function $\alpha$ may generate all possible $k!$ permutations of the $k$ points. We make a four-dimensional table ANS, which is indexed by the number of permutations (one dimension with $k!$ choices) as well as the possible positions of the predecessor (3 dimensions with $k+1$ choices for the $3$-sided range). Each cell of array ANS contains the position of the point with the maximum $x$-coordinate for a given permutation that corresponds to a tree $T^j_3$ and the $3$-sided range. Each tree $T^j_3$ corresponds to a permutation index that indexes one dimension of table ANS. The other $3$ indices are generated by $2$ predecessor queries on the $x$-coordinate and one predecessor query on the $y$ coordinate. The size of ANS is $O(k!(k+1)^3)$ and obviously it is common for all trees in the third layer of the MLR-tree. To build it, we proceed as follows:

We attach a unique label in the range $[1,k!]$ to each one of the $k!$ permutations corresponding to the respective index in array ANS. This label is constructed by enumerating systematically all $k!$ permutations and keeping them in an array of labels. Each label is represented by $O(k\log{k})$ bits. Each tree in the third layer is attached with such a label based on the permutation generated by the $y$-coordinates of its points. This is the only step in the building process that requires knowledge of the trees $T^j_3$. Then, we compute for every permutation and for every possible combination of the three predecessors the rank of the point with the maximum $x$-coordinate. This can be done by a single scan of the permutation for each possible combination of the predecessor queries. 
 
Although the skyline query changes to incorporate the division of the structure into $3$ layers, these changes are not extensive. Let $q=[a,b] \times(-\infty,d]$ be the initial range query. To answer this query on the three layered structure we access the layer $3$ trees containing $a$ and $b$ by using the $A'$ array. Then, we locate the subtrees $T^i_2$ and $T^j_2$ containing the representative leaves of the accessed layer $3$ trees. The roots of these subtrees are leaves of $T_1$. The MLR query algorithm described in~\ref{ssec:nonlinearmlr} is executed on $T_1$ with these leaves as arguments. Once we reach the node with the maximum $x$-coordinate, we continue in the layer $2$ tree corresponding to the representative with the maximum $x$-coordinate located in $T_1$. The same query algorithm is executed on this layer $2$ tree and then we move similarly to a tree $T^j_3$ in the third layer. We make three predecessor queries for $a$, $b$, and $d$ in $T^j_3$ and we use the ANS table to locate the point with the maximum $x$-coordinate by retrieving the permutation index of $T^j_3$. Let the point $z=(z_{x}, z_{y})$ be the desired point at the third layer. We go back to $T_1$. The range query now becomes $q=[a, z_x]\times (-\infty, z_y]$ and iterate as described in~\ref{ssec:nonlinearmlr}.

The total space required for the data structure depends on the size of each of the three layers. For the first layer, the MLR-tree on the $O\left(\frac{N}{\log^2{N}}\right)$ representatives requires $O\left(\frac{N}{\log^2{N}}\log^2{\left(\frac{N}{\log^2{N}}\right)}\right)=O(N)$ linear space for the leaf structures (all $P_{\ell}$ structures for each leaf $\ell$ are structured as $q^*$-heaps and h-RMQ structures requiring linear space). For the $S_v$ structures, the total space needed is $O\left(S_{PAM}\left(\frac{N}{\log^2{N}}\right)\log{N}\right)$. The second layer consists of $O\left(\frac{N}{\log^2{N}}\right)$ trees with $O\left(\left(\frac{log{N}}{\log{\log{N}}}\right)^2\right)$ representative points of the third layer each. Since each one of these trees is itself an MLR-tree its size is $O\left(\frac{\log^2{N}}{\log^2{\log^2{N}}}\log^2{\left(\frac{\log^2{N}}{\log^2{\log^2{N}}}\right)}\right)=O(\log^2{N})$. For the $S_v$ structures for each tree in the second layer we need $O\left(S_{PAM}\left(\frac{\log^2{N}}{\log^2{\log{N}}}\right)\log{\log{N}}\right)$ space. In total, the space for the second layer is $O\left(N+\frac{N\log{\log{N}}}{\log^2{N}}S_{PAM}\left(\frac{\log^2{N}}{\log^2{\log{N}}}\right)\right)$. In the third layer, we use linear space for the two predecessor data structures ($q^*$-heaps) as well as a table of size $O((4\log^2{\log{N}})!(4\log^2{\log{N}}+1)^3)$, which is $O(N)$. The construction time of the data structure can be similarly derived taking into account that the ANS table can be constructed in $O(N)$ time. As for the query, we get an $O(1)$ number of predecessor queries per iteration, in which iteration we report a point  on the skyline. The following theorem summarizes the result (note that $sort(N)$ is the time needed to sort a list of $N$ elements):\\

%\\ 
%\begin{theorem}
%The Lazy B-Tree supports the search operation in $O(\log_B N)$ worst-case block %transfers and update operations 
%in $O(1)$ worst-case block transfers, provided that the update position is given.
%\end{theorem}

\begin{theorem} \label{thm:static-linear-main-MLR}
Given a set of $N$ points on the 2-d grid $[1,M] \times [1,M]$, we can store them in a static main memory data structure that can be constructed in\\ 
$O\left(N\log{N}+M+C_{PAM}\left(\frac{N}{\log^2{N}}\right)\log{N}+C_{PAM}\left(\frac{\log^2{N}}{\log^2{\log{N}}}\right)\frac{N\log{\log{N}}}{\log^2{N}}\right)$ time using $O\Bigl(N+M+S_{PAM}\left(\frac{N}{\log^2{N}}\right)\log{N}+
S_{PAM}\left(\frac{\log^2{N}}{\log^2{\log{N}}}\right)\frac{N\log{\log{N}}}{\log^2{N}}\Bigr)$ space. It supports skyline queries in a $3$-sided range in $O(t \cdot t_{PAM}(N))$ worst-case time, where $t$ is the answer size.
\end{theorem}

\subsubsection{A Note on External Memory} \label{sssec:external}

The result can be easily extended to external memory as well.
The base tree is a static $B$-tree, where $B$ is the size of the block. One change to the structure is related to the definition of $P_{\ell,left}^{\tau}$ and $P_{\ell,right}^{\tau}$. In particular, $P_{\ell,left}^{\tau}$ (and similarly $P_{\ell,right}^{\tau}$) correspond to the node with the minimum $y$-coordinate among all nodes that are children of the nodes in $P_{\ell}^{\tau}$ and are to the left of a node $v$ in $P_{\ell}^{\tau}$ among all children of the father of $v$ that also belongs to $P_{\ell}^{\tau}$. This means that $P_{\ell,left}^{\tau}$ may contain $O(\log_B{N})$ nodes and each leaf $\ell$ may have $O(\log_B{N})$ such lists. Another change is related to the level $3$ trees. We make the assumption that $\log^2{\log{N}}=O(B)$, which means that a level $3$ tree can be easily stored in $O(1)$ blocks of size $B$ and as a result there is no need to use tabulation. See Figure~\ref{fig:linearmlr} for a depiction of the tree. To get a feeling of the problem size that would violate this assumption, we get that $\log^2{\log{N}}\leq B$ when $N\leq 2^{2^{\sqrt{B}}}$, which even for small values of block size, like $B=256$, we get that a level $3$ tree can be stored in a block when $N\leq 2^{2^{16}}$, which is a number much larger than a googol ($10^{100}$). The changes in the query algorithm are insignificant and mainly related to the change of the definition of $P_{\ell,left}^{\tau}$ and $P_{\ell,right}^{\tau}$ for all $\ell$ and ${\tau}$.

The following theorem is an easy extension of Theorem~\ref{thm:static-linear-main-MLR} for external memory.

\begin{theorem} \label{thm:static-linear-main-MLR-external}
Given a set of $N$ points on the 2-d grid $[1,M] \times [1,M]$, we can store them in a static external memory data structure that can be constructed in \\
$O\left(sort(N)+\frac{M}{B}+C_{PAM}\left(\frac{N}{\log^2{N}}\right)\log_B{N}+C_{PAM}\left(\frac{\log^2{N}}{\log^2{\log{N}}}\right)\frac{N\log_B{\log{N}}}{\log^2{N}}\right)$ using \\
$O\left(\frac{N+M}{B}+S_{PAM}\left(\frac{N}{\log^2{N}}\right)\log_B{N}+S_{PAM}\left(\frac{\log^2{N}}{\log^2{\log{N}}}\right)\frac{N\log_B{\log{N}}}{\log^2{N}}\right)$ space. It supports skyline queries in a $3$-sided range in $O\left((t/B) \cdot t_{PAM}\left(N\right)\right)$ I/Os, where $t$ is the answer size.
\end{theorem}

\subsubsection{Results for the Static Case in Main Memory and External Memory} \label{sssec:results}

Applying Theorem~\ref{thm:static-linear-main-MLR} for various implementation of PAMs in main memory we get different results that are summarized in the following:

\begin{itemize}
\item \textbf{Binary Trees:} Assuming that the PAM is a simple binary tree, the MLR tree uses $O(N+M)$ space, can be constructed in $O(N\log{N}+M)$ time (by merging the sorted lists in linear time in a new sorted list with respect to the $y$-coordinate) and has a query time of $(t\log{N})$.

\item \textbf{van Embde Boas trees~\cite{vanEmdeBoas1976}:} Assuming that the PAM is a van Emde Boas tree, the MLR tree uses $O(N+M)$ space, can be constructed in $O(N\log{N}+M)$ time and has a query time of $(t\log{\log{N}})$.

\item \textbf{IS-tree~\cite{KMSTTZ06} (Random Input):} Assuming that the PAM is an Interpolation Search Tree and that the elements are drawn from a $\left(\frac{N}{g},\ln^{O(1)}{N}\right)$-smooth distribution, where $g$ is a constant, then the MLR tree uses $O(N+M)$ space, can be constructed in $O(N\log{N}+M)$ time and has an expected query time of $O(t)$.

\item \textbf{RSA~\cite{BKS11} (Random Input):} Assume that the PAM is the Random Search Array and that the elements are drawn from a $(N^{\gamma},N^{\alpha})$-input distribution, where $1<\alpha<1$ and $\gamma>0$ (vastly larger than the family of distributions for the IS-tree). The MLR tree uses $O(N^{1+\delta}+M)$ space, can be constructed in $O(N^{1+\delta}+M)$ time and has an expected query time of $O(t)$ with high probability, where $\delta>0$ is an arbitrarily chosen constant.

\item \textbf{BPQ~\cite{MKYN05} (Batched Predecessor Queries):} Assume that the PAM is the Data Structure presented in \cite{MKYN05}, that answers batched predecessor queries in $O(1)$ amortized time. In particular, supports $p(M)$ queries in $O(\frac{loglogM}{p(M)})$ time per query and requires $O(N^{p(M)})$ space for any  $p(M)=loglogM$. In this case, the MLR-tree uses exponential space and supports batched skyline queries in optimal $O(t)$ amortized time.

\end{itemize}

Similarly, applying Theorem~\ref{thm:static-linear-main-MLR-external} for various implementations of PAMs in external memory we get the following results.

\begin{itemize}
\item \textbf{$B$-trees:} Assuming that the PAM is a simple binary tree, the MLR tree uses $O\left(\frac{N+M}{B}\right)$ space, can be constructed in $O\left(sort(N)+\frac{M}{B}\right)$ time and has a query time of $(t\log_B{N})$.

\item \textbf{$ISB$-trees:} Assuming that the PAM is an $ISB$-tree~\cite{KMMSTTZ05} for discrete distributions as indicated by \cite{KMSTTZ06}and assuming that the coordinates of the points are generated by a smooth discrete distribution for each dimension independently, the MLR tree uses $O\left(\frac{N+M}{B}\right)$ space, can be constructed in $O\left(sort(N)+\frac{M}{B}\right)$ time and has a query time of 
$((t/B)\log_B{\log{N}})$. For a smaller set of distributions, the query time can be reduced to $O(t/B)$ (see~\cite{KMSTTZ06}).
\end{itemize}

\subsubsection{The \texorpdfstring{$3$}{3}-sided Skyline Problem is at least as hard as the Predecessor Problem} \label{sssec:eq_pred}

Our approach makes explicit that the main bottleneck in the $3$-sided skyline problem is the predecessor problem. At this point we show that this is not an artifact of our approach but in fact the $3$-sided skyline problem is at least as much difficult as the predecessor problem. This means that we can only hope for bounds which resemble the bounds in the predecessor problem and not better than these. In the following, we show how the predecessor problem can be solved efficiently by the $3$-sided skyline problem implying that the same lower bounds with the predecessor problem apply. Note that this is folklore knowledge and we provide it here for the sake of completeness as well as because our approach is explicitly heavily dependent on the predecessor problem. 

Assume a sorted sequence $A=x_1,x_2,\ldots,x_N$ of $N$ integer elements chosen from the range $[1,M]$. We construct in $O(N)$ time a set of points $S=\{p_1=(x_1,x_1),p_2=(x_2,x_2),\ldots,p_N=(x_N,x_N)\}$ in two dimensions. Assume that we use MAX-MAX semantics. Let a predecessor query $pred(a)$ on $A$, where $a\in [1,M]$ an arbitrary integer and let the answer of the query be $x_i, 1\leq i\leq N$. We make the $3$-sided skyline query with range $[1,M]\times(-\infty,a]$. This means that there is no restriction on the $x$-coordinate and we only wish to find the skyline of all points that have $y$-coordinate $\leq a$. 

We argue that the result of this particular $3$-sided skyline query is point $p_i=(x_i,x_i)$. Indeed, by construction, each point $p_i$ dominates all points $p_j$, such that $j<i$, which means that for all $3$-sided ranges the skyline consists of at most one point. Since all points satisfy the restriction on the $x$-coordinate, we must consider only the points with $y$-coordinate $\leq a$. The point with the largest $x$-coordinate and with $y$-coordinate $\leq a$ is $p_i=(x_i,x_i)$. This point dominates all other points and as a result it is the only point on the particular skyline. As a result, we have answered the predecessor query as well. Since our approach has similar bounds with those optimal bounds of the predecessor problem, we can state that our solutions are optimal.

\section{The Dynamic MLR-tree} \label{ssec:dynamiclinearMLR}

Making dynamic the layered MLR tree described in~\ref{ssec:staticlinearMLR} involves all layers. The following issues must be tackled in order to make the MLR-tree dynamic: 1. use of a dynamic tree structure with care to how rebalancing operations are performed, 2. the layer $3$ trees must have variable size within a predefined range, rebuilding them appropriately as soon as they violate this bound (by splitting or merging/sharing with adjacent trees) - similarly, the permutation index must be appropriately defined in order to allow for variable length permutations and 3. all arrays attached to nodes or leaves as well as array $A'$ must be updated efficiently.

To begin with, global rebuilding~\cite{DBLP:books/sp/Overmars83} is used in order to maintain the structure. In particular, let $N_0$ be the number of elements stored at the time of the latest reconstruction. After that time when the number of updates exceeds $rN_0$, where $0<r<1$ is a constant, then the whole data structure is reconstructed taking into account that the number of elements is $rN_0$. In this way, it is guaranteed that the current number of elements $N$ is always within the range $[(1-r)N_0,(1+r)N_0]$. We call the time between two successive reconstructions an {\em epoch}. The tree structure used for the first two layers is a weight-balanced tree, like the $BB[a]$-trees~\cite{DBLP:journals/siamcomp/NievergeltR73} or the weight-balanced $(a,b)$-trees~\cite{DBLP:journals/siamcomp/ArgeV03}. In the latter case, the tree is not binary and the definition of lists $L^{\tau}_{\ell}$ and $R^{\tau}_{\ell}$ is extended analogously to external memory static MLR-tree in order to take into account the appropriate nodes. 

Henceforth, assume for brevity that $k=\log^2{\log{N}}$. We impose that all trees at layer $3$ will have size within the range $[k/4,k]$. To compute the permutation index, if the size of the layer $3$ tree is $<k$, then we pad the increasing sequence of elements in the tree with $+\infty$ values in order to have exactly size $k$ (alternatively, we could count also the number of subsets of size in the range $[k/4,k]$ increasing the size of the table ANS but not exceeding the $O(N)$ bound). In addition, the array $A'$ that indexes the leaves is structured with a PAM since it must be dynamic as well.

%The update algorithms

Assume that an update operation takes place. The following discussion concerns the case of inserting a new point $q$ since the case of deleting an existing point $q$ from the structure is symmetric. First, $A'$ is used to locate the predecessor of $q_x$, and in particular to locate the tree $T^j_3$ of layer $3$ that contains the predecessor of $q_x$. Array $A'$ is updated accordingly. The predecessor of $q_x$ in $T^j_3$ is located by using the respective $q^*$-heap. If $|T^j_3|\in [k/4,k]$, then $q_x$ and $q_y$ are inserted in the respective $q^*$-heaps. If $|T^j_3|>k$, then $T^j_3$ is split into two trees with size approximately $\frac{k}{2}$. This means that $4$ new $q^*$-heaps must be constructed while two new permutation indices must be computed for the two new trees. Let $T^i_2$ be the layer $2$ tree that gets the new leaf. Note that $T^i_2$ is affected either structurally, when one of its leaves $\ell$ at layer $3$ splits as in this case ($\ell$ is $T^j_3$) or it is affected without structural changes, when $q_y$ is minimum among all the $y$-coordinates of $T^j_3$ and thus the representative of $T^j_3$ changes. In the latter case, all structures $S_v$ on the path $P_{\ell}$ of $T^i_2$ must be updated with the new point. In addition, let $v$ be the highest node with height $h_v$ in $T^i_2$ that has $p_v=q$ (the point with the minimum $y$-coordinate in its subtree changes to $q$). Then, for all leaves $\ell$ in the subtree of $v$, the $q^*$-heaps for $L^{\tau}_{\ell}$ and $R^{\tau}_{\ell}$ as well as the h-RMQ structures are updated, given that ${\tau}\geq h_v$. In the former case, we make rebalancing operations on the internal nodes of $T^i_2$ on the path $P_{\ell}$. These rebalancing operations result in changing as in the previous case the $q^*$-heaps for the $L^{\tau}_{\ell}$ and $R^{\tau}_{\ell}$ while the respective $S_v$ structures of the node $v$ that is rebalanced have to be recomputed. Similar changes happen to the tree $T_1$ of the first layer given that either a tree of the second layer splits or its minimum element is updated. In case of deleting $x$, the $3$ layers of the MLR-tree are handled similarly.

%Complexity

In the following discussion assume that the time complexity of the update operation supported by the PAM on $N$ elements is $O(U_{PAM}(N))$. The change of the point with the minimum $y$-coordinate can always propagate from $T^j_3$ to the root of $T_1$. $T^j_3$ can be updated in $O(|T^j_3|)$ time since the two updates in $q^*$-heaps cost $O(1)$ while the computation of the permutation index costs $O(|T^j_3|)$. Let the respective tree in the second layer be $T^i_2$. Then, the cost for changing the point with the minimum $y$-coordinate in each node on the path from the leaf to the root of $T^i_2$ is related to the update cost for the $L^{\tau}_{\ell}$ and $R^{\tau}_{\ell}$ lists as well as for the $S_v$ structures. In particular, all $O(|T^i_2|\log{|T^i_2|})$ lists $L^{\tau}_{\ell}$ and $R^{\tau}_{\ell}$ are updated (deletion of the previous point and insertion of the new one in a $q^*$-heap) in $O(|T^i_2|\log{|T^i_2|})$ time. Similarly, a deletion and an insertion is carried out in each $S_v$ structure in $O(U_{PAM}(|T^i_2|)\log{|T^i_2|})$ total time. The same holds for the tree $T_1$ getting a total complexity of $O((|T_1|+U_{PAM}(|T_1|))\log{|T_1|})$. 

Rebalancing operations on the level $2$ trees as well as on the level $1$ tree of the structure may be applied when splits or fusions of leaves of level $2$ trees take place. Since level $2$ trees are exponentially smaller than the level $1$ tree and they are the same, the cost is dominated by the rebalancing operations at $T_1$. Assume an update operation at a leaf $\ell$ of $T_1$. In the worst case, each $S_v$ structure may have to be reconstructed and similarly to the previous paragraph the $L^{\tau}_{\ell}$ and $R^{\tau}_{\ell}$ structures need to be updated. The total cost is equal to $O(|T_1|\log^2{|T_1|})$ for the $O(|T_1|\log{|T_1|})$ lists while it is $O(C_{PAM}(|T_1|))$ for the $S_v$ structures since the reconstruction of the $S$ structure of the root dominates the cost. One can similarly reason for level $2$ trees. However, the amortized cost is way lower for two reasons: 1. There is an update at a leaf of $T_1$ roughly every $O(log^2{N})$ update operations and 2. The weight property of the tree structures guarantees that costly operations are rare. By using a standard weight property argument combined with the above two reasons we get that the amortized rebalancing cost is $O\left(\log^2{N}+\frac{C_{PAM}(|T_1|)}{N}\log{N}\right)$. This amortized cost is dominated by the cost to update the minimum element, in which case the worst-case as well as the amortized case coincide.The following theorem summarizes the result:

\begin{theorem} \label{thm:MLRupdate}
Given a set of $N$ points on the 2-d grid $[1,M] \times [1,M]$, we can store them in a dynamic main memory data structure that uses $O\Bigl(N+M+S_{PAM}\left(\frac{N}{\log^2{N}}\right)\log{N}+$ \\ $S_{PAM}\left(\frac{\log^2{N}}{\log^2{\log{N}}}\right)\frac{N\log{\log{N}}}{\log^2{N}}\Bigr)$ space and supports update operations in \\
$O\left(C_{PAM}\left(\frac{N}{\log^2{N}}\right)+ \frac{N}{\log^2{N}}\log{\left(\frac{N}{\log^2{N}}\right)} \right)$ in the worst case. It supports skyline queries in a $3$-sided range in $O(t \cdot t_{PAM}(N))$ worst-case time, where $t$ is the answer size.
\end{theorem}

The inefficiency of the update operations is overwhelming. Although rebalancing operations are efficient in an amortized sense, the change of minimum depends on the user and in principle this change can propagate to the root in each update operation. In the following, we overcome this problem by making a rather strong assumption about the distribution of the points.

\subsection{Exploiting the Distribution of the Elements}

To reduce the huge worst-case update cost of Theorem~\ref{thm:MLRupdate} we have to tackle the propagation of minimum elements. Assume that a new point $q=(q_x,q_y)$ is to be inserted in the MLR-tree. Let $q$ be stored in level $2$ tree $T^i_2$ according to $q_x$. We call the point $q$ {\em violating} if $q_y$ is the minimum $y$-coordinate among all $y$-coordinates of the points in $T^i_2$. When a new point is violating it means that an update operation must be performed on $T_1$. In the following, we show that under assumptions on the generating distributions of the $x$ and $y$ coordinates of points we prove that during an epoch~\footnote{Recall than an epoch is the time between two successive reconstructions of the structure defined by the $\Theta(N)$ update operations.} only $O(\log{N})$ violations will happen. We provide a sketch of the structure since it is an easy adaptation of the probabilistic results of~\cite{KMSTTZ06,KPSTT10}. 

We assume that all points have their $x$ coordinate generated by the same discrete distribution $\mu$ that is $(f_1(N)=N/(\log \log N)^{1+\epsilon}, f_2(N)=N^{1-\delta})$-smooth, where $\epsilon > 0$ and $\delta \in (0,1)$ are constants. We also assume that the $y$ coordinates of all points are generated by a restricted set of discrete distributions $\mathcal{Y}$, independently of the distribution of the $x$ coordinate. We later show the properties that $Y$ must have and provide specific examples. Finally, we assume that deletions are equiprobable for each existing point in the structure. In a nutshell, the structure requires that during an epoch tree $T_1$ remains intact and only level $2$ and level $3$ trees are updated. All violating points are stored explicitly and since they are only a few during an epoch, we can easily support the query operation. After the end of the epoch, the new structure has no violating points stored explicitly. 

The construction of the static tree $T_1$ now follows the lines of~\cite{KMSTTZ06}. Without going into details, assume that the $x$ coordinates are in the range $[x_1,x_2]$. Then, this range is recursively divided into $f_1(N)$ subranges. The terminating condition for the recursion is when a subrange has $\leq \log^2{N}$ elements. Note that the bounds of these subranges do not depend on the stored elements but only on the properties of the distribution. This construction is necessary to ensure certain probabilistic properties for discrete distributions. However, instead of building an interpolation search tree, we build a binary tree on these subranges and then continue building the lists of the leaves and the internal nodes as in the previous structures. The elements within each subrange correspond to a level $2$ tree whose leaves are level $3$ trees. Theorem~1 and Lemma~2 of~\cite{KMSTTZ06} imply the following theorem with respect to each epoch:

\begin{theorem}\label{thm:prob_constr_height}
The construction of the terminating subranges defining the level $2$ trees can be performed in $O(N)$ time in expectation with high probability. Each level $2$ tree has $\Theta(\log^2{N})$ points in expectation with high probability during an epoch.
\end{theorem} 

The above theorem guarantees that the size of the buckets is not expected to change considerably and as a result we are allowed to assume that no update operations will happen on $T_1$. This is the result of assuming that the $x$ coordinates of the points inserted are generated by an $(N/(\log \log N)^{1+\epsilon}, N^{1-\delta})$-smooth distribution.

The reduction of the number of violating points during an epoch is taken care of by our assumption that the $y$ coordinates follow a distribution that belongs to the $\mathcal{Y}$ family of distributions. Recall that the $y$ coordinate is drawn from the integers in the range $[1,M]$ according to a distribution $\mu'\in \mathcal{Y}$. Let an arbitrary point $p=(p_x,p_y)$ and let $\alpha=Pr[p_y>1]$ (probability that the $y$ coordinate is strictly larger than the minimum element in $[1,M]$). Then, by slightly altering Theorem 3.4 in~\cite{KPSTT10} (to accommodate discrete distributions) we get the following:

\begin{theorem} \label{thm:insdel:n}
For a sequence of $\Theta(n)$ updates, the expected number of violations is $O(\log{n})$, assuming that $x$ coordinates are drawn from an $(N/(\log \log N)^{1+\epsilon},N^{1-\delta})$-smooth distribution, where $\epsilon > 0$ and $\delta \in (0,1)$ are constants, and the $y$ coordinates are drawn from the restricted class of distributions $\mathcal{Y}$ such that it holds that $\alpha\leq \left(\frac{\log{N}}{N} \right)^{\frac{1}{\log{N}}}\rightarrow e^{-1}$, where $\alpha=Pr[p_y>1]$ for an arbitrary point $p=(p_x,p_y)$.
\end{theorem}

\begin{figure}[!t]
	\centering
		\includegraphics[width=1.0\textwidth]{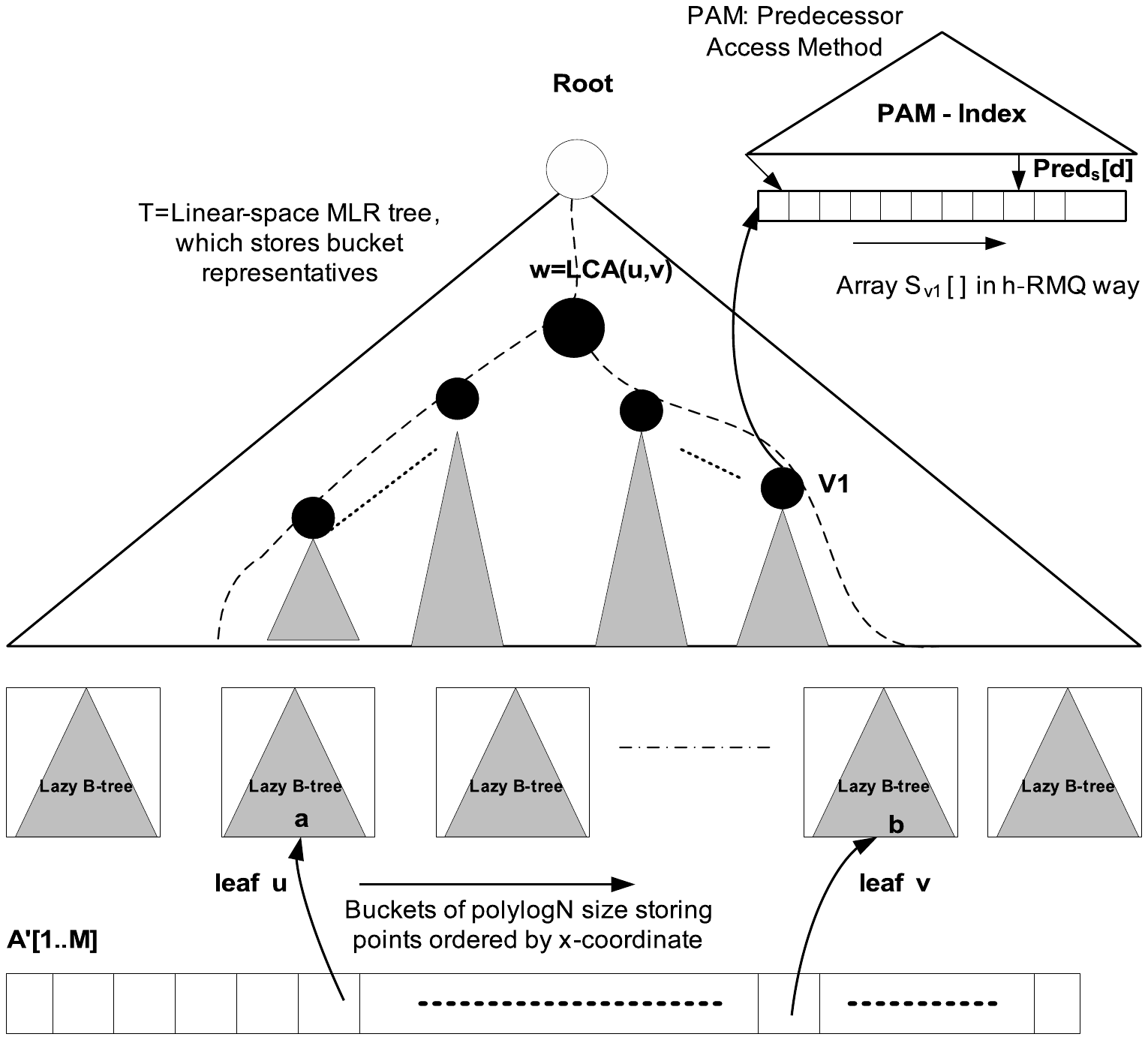}
		\vspace{-2cm}
	\caption{The dynamic linear-space MLR-tree.}
	\label{fig:dynamicmlr}
\end{figure}

The {\em Power Law} and {\em Zipfian} distributions have the aforementioned property that $q \leq \left(\frac{\log{N}}{N} \right)^{(\log{N})^{-1}} \rightarrow e^{-1}$ as $n \rightarrow \infty$.

The theorem for 3-sided dynamic skyline queries follows:

\begin{theorem} \label{thm:MLRupdate-prob}
Given a set of $N$ points on the 2-d grid $[1,M] \times [1,M]$, whose $x$ coordinates are generated by an $(N/(\log \log N)^{1+\epsilon},N^{1-\delta})$-smooth distribution, where $\epsilon > 0$ and $\delta \in (0,1)$ are constants, and the $y$ coordinates are drawn from the restricted class of distributions $\mathcal{Y}$ that contain the power law and zipfian distributions, we can store them in a dynamic main memory data structure that uses $O\left(N+S_{PAM}\left(\frac{N}{\log^2{N}}\right)\log{N}+S_{PAM}\left(\frac{\log^2{N}}{\log^2{\log{N}}}\right)\frac{N\log{\log{N}}}{\log^2{N}}\right)$ space and supports update operations in $O\left(C_{PAM}\left(\frac{\log^2{N}}{\log^2{\log{N}}}\right)+ \frac{\log^2{N}}{\log^2{\log{N}}}\log{\left(\frac{\log^2{N}}{\log^2{\log{N}}}\right)} \right)$ in expectation with high probability. It supports skyline queries in a $3$-sided range in $O(t \cdot t_{PAM}(N))$ worst-case time, where $t$ is the answer size.
\end{theorem}

Similarly to Section~\ref{sssec:results} we can employ various PAM structures and get different results. For example, by employing the IS-tree as described in Section~\ref{sssec:results}, the structure uses $O(N)$ space, supports updates in $O(\log^2{N}\log{\log{N}})$ time in expectation with high probability and has an expected query time of $O(t)$.

\section{Conclusions and Future Work} \label{sec:conclusions}

In this paper we presented the MLR (Modified Layered Range) tree-structure providing an optimal time solution for finding planar skyline points in a 3-sided orthogonal range in both RAM and I/O model on the grid $[1,M] \times [1,M]$, by single scanning not all the sorted points but the points of the answer $SKY(P)$ only. 
Currently, we are working towards an experimental performance evaluation towards comparing our approach with existing approaches for skyline computation such as BBS \cite{PTFS05}. This study will bring forward the nice properties of the MLR approach since not only offers non-trivial expected performance guarantees but also works very well in practice.

Another interest research direction involves the transformation of the MLR-tree into a cache-oblivious \cite{ABDMM07,BDC05,BBFGHHIO11} data structure, so as to be independent  on the number of memory levels, the block sizes and number of blocks at each level, or the relative speeds of memory access. Typically, a cache-oblivious algorithm works by a recursive divide and conquer algorithm, where the problem 
is divided into smaller and smaller subproblems. Eventually, one reaches a subproblem size that fits into cache, regardless of the cache size. The MLR-tree consists of arrays L, R, P that could be trivially transformed in cache-oblivious model, 
a number o $k$ recursive layers where the last layer (microtree) could be fit into cache (disk block) regardless of the cache size. 
Moreover, MLR-tree uses auxiliary data structures for predecessor and h-RMQ queries that have been already implemented in the cache-oblivious model (see \cite{BR12} and \cite{HMR10} respectively). 
In addition, the improvement of update performance of MLR-tree constitutes a challenging open problem. A bucketing technique (see figure\ref{fig:dynamicmlr}), where the upper level MLR-tree stores bucket representatives
and the lower level is a Lazy B-tree that supports updates with $O(1)$ rebalancing operations, could improve more the total update performance.

%%
%% Bibliography
%%

%% Either use bibtex (recommended), 

%\bibliography{lipics-v2016-sample-article}

%% .. or use the thebibliography environment explicitely

\end{document}